\documentclass[pre,twocolumn,showpacs,superscriptaddress]{revtex4-1}
\usepackage{graphicx}
\usepackage{amsmath,amssymb}
\usepackage{upgreek}
\newcommand{\kB}{k_{\mathrm{B}}}
\newcommand{\kT}{\kB T}

\newcommand{\um}{\upmu\mathrm{m}}
\newcommand{\umpersec}{\um\,\mathrm{s}^{-1}}
\newcommand{\umsqpersec}{\um^2\,\mathrm{s}^{-1}}
  
\newcommand{\Pe}{{\cal P}}
\newcommand{\PeCDD}{\Pe_{\mathrm{coll}}}
\newcommand{\Pefilm}{\Pe_{\mathrm{film}}}

\newcommand{\vev}{v_{\mathrm{ev}}}
\newcommand{\phip}{\phi_{\mathrm{p}}}
\newcommand{\phipinit}{\phi_{0}}

\newcommand{\colloid}{{c}}
\newcommand{\polymer}{{p}}
\newcommand{\solvent}{{s}}

\newcommand{\Dc}{D_\colloid}
\newcommand{\Rc}{R_\colloid}
\newcommand{\Dp}{D_\polymer}
\newcommand{\rhop}{\rho_\polymer}
\newcommand{\vs}{v_\solvent}
\newcommand{\vc}{v_\colloid}
\newcommand{\zint}{z_{\mathrm{int}}}
\newcommand{\tstar}{t^*}
\newcommand{\Lstar}{L^*}

\newcommand{\Peclet}{P\'eclet}

\newcommand{\latin}[1]{{\itshape #1}}
\newcommand{\eg}{\latin{e.\,g.}}
\newcommand{\ie}{\latin{i.\,e.}}
\newcommand{\etal}{\latin{et al.}}
\newcommand{\etc}{\latin{etc}}
\newcommand{\via}{\latin{via}}
\newcommand{\viz}{\latin{viz.}}
\newcommand{\versus}{\latin{versus}}
\newcommand{\adhoc}{\latin{ad hoc}}

\newcommand{\french}[1]{{\itshape #1}}
\newcommand{\tourdeforce}{\french{tour-de-force}}

\DeclareMathOperator\erfc{erfc}
\newcommand{\Jc}{\Jvec_c}
\newcommand{\Jp}{\Jvec_p}

\newcommand{\dvol}{{\mathrm d}\rvec}
\newcommand{\Vcp}{V_{cp}}
\newcommand{\myvec}[1]{\mathbf{#1}}
\newcommand{\Jvec}{\myvec{J}}
\newcommand{\Uvec}{\myvec{U}}
\newcommand{\rvec}{\myvec{r}}

\newcommand{\german}[1]{{\itshape #1}}
\newcommand{\ansatz}{\german{ansatz}}

\newcommand{\Eqref}[1]{Eq.~\eqref{#1}}

\newcommand{\Refcite}[1]{Ref.~\onlinecite{#1}}

\newcommand{\Figref}[1]{Fig.~\ref{#1}}

\begin{document}

\title{Diffusiophoresis in non-adsorbing polymer solutions:\\
  the Asakura-Oosawa model and stratification in drying films}

\author{Richard P. Sear}

\affiliation{Department of Physics, University of Surrey, Guildford,
  GU2 7XH, UK}

\email{r.sear@surrey.ac.uk}

\author{Patrick B. Warren}

\affiliation{Unilever R\&D Port Sunlight, Quarry Road East, Bebington,
  Wirral, CH63 3JW, UK}

\email{patrick.warren@unilever.com}

\begin{abstract}
  A colloidal particle placed in an inhomogeneous solution of
  smaller non-adsorbing polymers will move towards regions of lower
  polymer concentration, in order to reduce the free energy of the
  interface between the surface of the particle and the solution.
  This phenomenon is known as diffusiophoresis. Treating the polymer
  as penetrable hard spheres, as in the Asakura-Oosawa model, a simple
  analytic expression for the diffusiophoretic drift velocity can be
  obtained. In the context of drying films we show that
  diffusiophoresis by this mechanism can lead to
  stratification under easily accessible experimental conditions.
  By stratification we mean
  spontaneous formation of
  a layer of polymer on top of a layer
  of the colloid.
  Transposed to the case of binary colloidal mixtures, this
  offers an explanation for the stratification observed recently in
  these systems [A. Fortini \etal, {Phys.\ Rev.\ Lett.} {\bf 116},
    118301 (2016)].  Our results emphasise the importance of treating
  solvent dynamics explicitly in these problems, and caution against
  the neglect of hydrodynamic interactions or the use of implicit
  solvent models in which the absence of solvent backflow results in
  an unbalanced osmotic force which gives rise to large but
  unphysical effects.
\end{abstract}

\maketitle

\section{Introduction}\label{sec:intro}
\begin{figure}[b]
  \includegraphics[width=6.5cm]{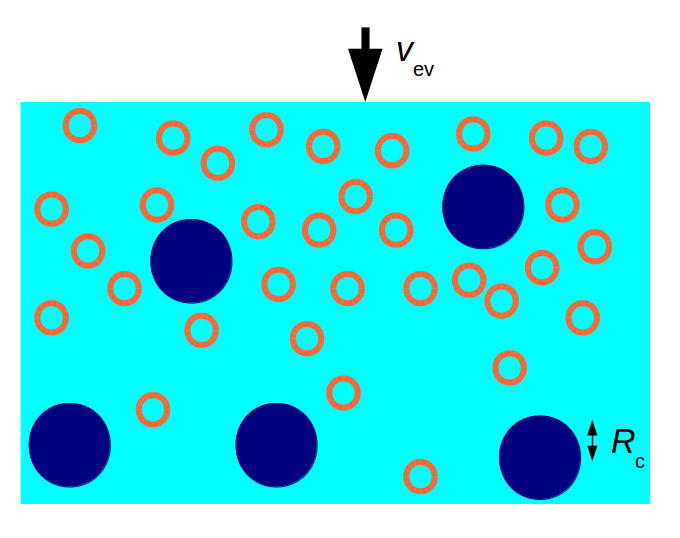}
  \caption{A schematic of a drying liquid (cyan) film containing
    colloidal particles (dark blue) and smaller polymer molecules
    (orange).  The evaporation speed is $\vev$.\label{film_schem}}
\end{figure}

Many coatings, from paints to cosmetics, form by the drying of a thin,
initially liquid, film.  The liquid film contains a dispersion of
colloidal particles and other non-volatile species that are left
behind after the liquid evaporates, 
and these form a solid film
\cite{Keddie:2010ta, routh13review}.  Due to the importance of
coatings made from drying colloidal suspensions, there is an extensive
literature on this process \cite{Antonietti:2000fl, Routh:2004jz,
  Reyes:2005dt, Reyes:2007ku, Luo:2008jd, Ekanayake:2009bs,
  Cardinal:2010be, Nikiforow:2010bi, trueman12, Atmuri:2012ks,
  Gorce:2014dl, Gromer:2015ko, Nunes2014}.  As the solvent (usually
water) evaporates, the liquid/air interface descends at some speed
$\vev$, pushing the non-volatile species such as colloidal particles
and polymer molecules, ahead of it (\Figref{film_schem}).

Drying is a non-equilibrium process, and it creates concentration
gradients. The concentration of colloid and polymer particles is high
in an accumulation zone just below the descending interface, and lower
near the bottom of the film. When there are concentration gradients in
a mixture, there will be diffusiophoresis. Diffusiophoresis is the
motion of one species in response to a gradient in the concentration
of another. In this work, we focus on the motion of colloidal
particles in response to a gradient in the concentration of smaller
polymer molecules.  We find that this diffusiophoretic motion can be
strong enough to exclude the colloidal particles from a top layer of
the drying film, \ie\ it can drive stratification into a layer of
small polymer molecules on top of a layer of the larger colloid
particles.

Recent experimental work \cite{fortini16, martin16, makepeace17} on making solid
films via the evaporation of mixtures
of colloidal particles,
has
found stratified dry films.
Here we argue that diffusiophoresis
provides a potential explanation for how this stratification occurs
spontaneously during drying.  

There has also been recent computer
simulation and modelling work on this phenomena \cite{fortini16,
  HNP17a, zhou17, fortini17, makepeace17, HNP17b}.  Indeed
stratification was first discovered in simulations by Fortini {\it et al.}
\cite{fortini16}.  However, to our knowledge nearly all of the
existing work used an implicit solvent, \ie\ modelled or simulated the
particles as diffusing in a uniform continuum of some viscosity
$\eta$, neglecting hydrodynamic interactions (a notable exception is
Cheng and Grest \cite{Cheng:2013cp} who use molecular dynamics with an
explicit solvent).  In models with an implicit solvent that is set to be
uniform, only gradients
in the chemical potentials
of the colloidal particles,
and hence in the osmotic pressure,
are possible. It is the gradient
in the chemical potential of the small
colloidal species,
or equivalently in the osmotic
pressure, that drives
stratification in these models
\cite{fortini16, HNP17a, zhou17}.  Here we consider the solvent
explicitly and find that in an evaporating film the solvent
\emph{backflow} leads to a counter-gradient in the solvent pressure,
that we expect to balance the gradient in the osmotic
pressure. Neglecting backflow, as has been done for simplicity in much
of the current work, therefore appears to be an unjustified approximation.

The current resurgence of interest in diffusiophoresis has largely
focused on particles driven by electrolyte gradients, where there are
significant effects additionally arising from diffuse liquid junction
potentials \cite{PAE+84, EAP88, SQ89a, ACY+08, Pri08, PCY+12, RSP13,
  FMH+14, KCR+15, PAN+15, BWA+16, MFW+16, SUS+16}.  However,
diffusiophoresis also occurs in gradients of uncharged solutes
\cite{ALP82, SQ89b, And89, VGG+16, MYB17, YMB17}, and we will apply
expressions from this literature to polymer solutions. Our model for a
polymer is the Asakura-Oosawa (AO) model \cite{asakura54, vrij76,
  BVS14}, where the polymer itself is ideal (no polymer-polymer
interactions) but is excluded from a layer of solvent of thickness $R$
around each colloidal particle.  The radius $R$ is comparable to the
radius of gyration of the polymer.  This is a very simple model, and
we believe that this paper can also serve as a pedagogical
introduction to diffusiophoresis, identifying the molecular origins of
the phenomena and highlighting the importance of solvent backflow
\cite{derjaguin47, ruckenstein81, And86, And89, MYB17}.

The remainder of this paper is divided into four sections.  In the
first section we 
obtain an expression for the
diffusiophoretic drift velocity of a large colloidal particle in a solution
of much smaller AO-model polymers.  Then we compare our approach in which
solvent flow is taken into account, with earlier work with an implicit
solvent, where solvent flow was neglected.  We then study
diffusiophoresis in a drying film containing colloidal particles and
smaller polymer molecules.  Our final section is a conclusion.

\begin{figure}
  \includegraphics[width=7.5cm]{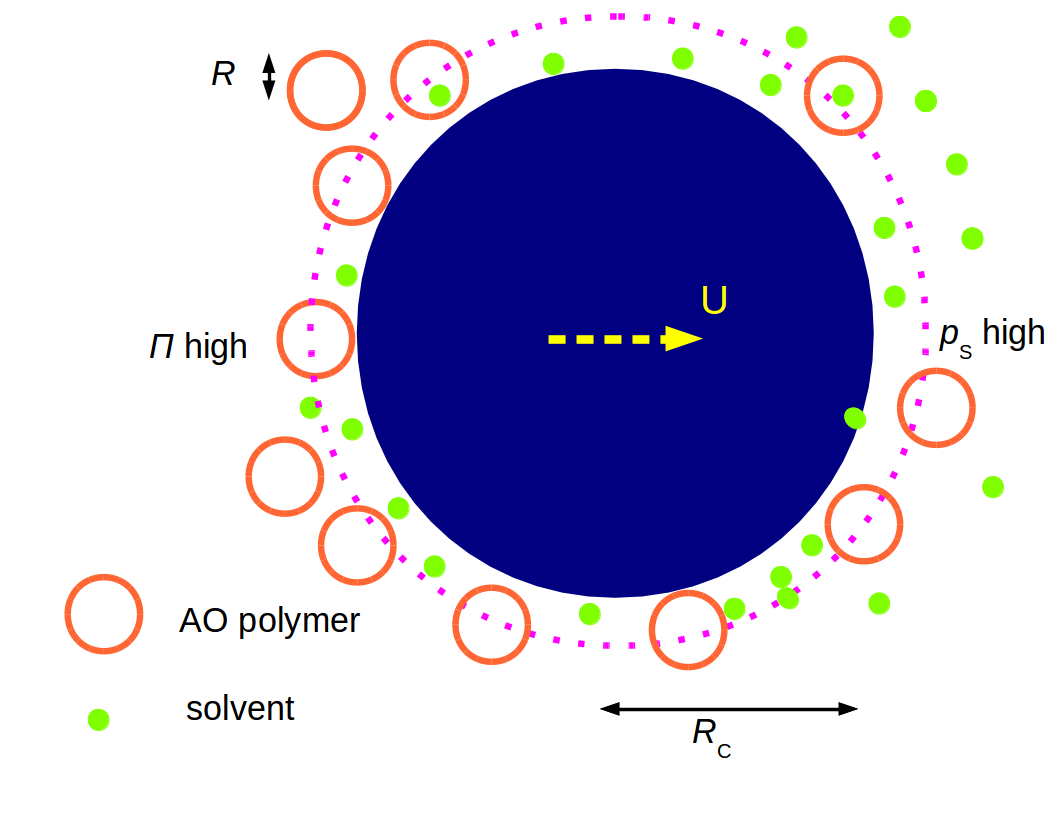}
  \caption{A schematic of a large particle (blue) immersed in a
    solution of AO-model polymer (orange) of radius $R$, in a solvent of
    smaller molecules (green).  The particle excludes the polymer from
    layer of width $R$ (indicated by a dotted line).
    Along the horizontal direction, there are
    gradients of both the polymer contribution 
    to the pressure ($\Pi$),
    and the solvent contribution ($p_{\mathrm{s}}$).\label{part_schem}}
\end{figure}

\section{Diffusiophoresis in the Asakura-Oosawa model}\label{sec:dp}
We are interested in determining the diffusiophoretic drift velocity $U$ of
a colloidal particle in a solution of much smaller polymer. There is a
gradient $\nabla \rhop$ in the concentration of the polymer.  This is
illustrated in \Figref{part_schem}.

\subsection{Slip velocity at flat hard wall}
\begin{figure}
  \includegraphics[width=6.5cm]{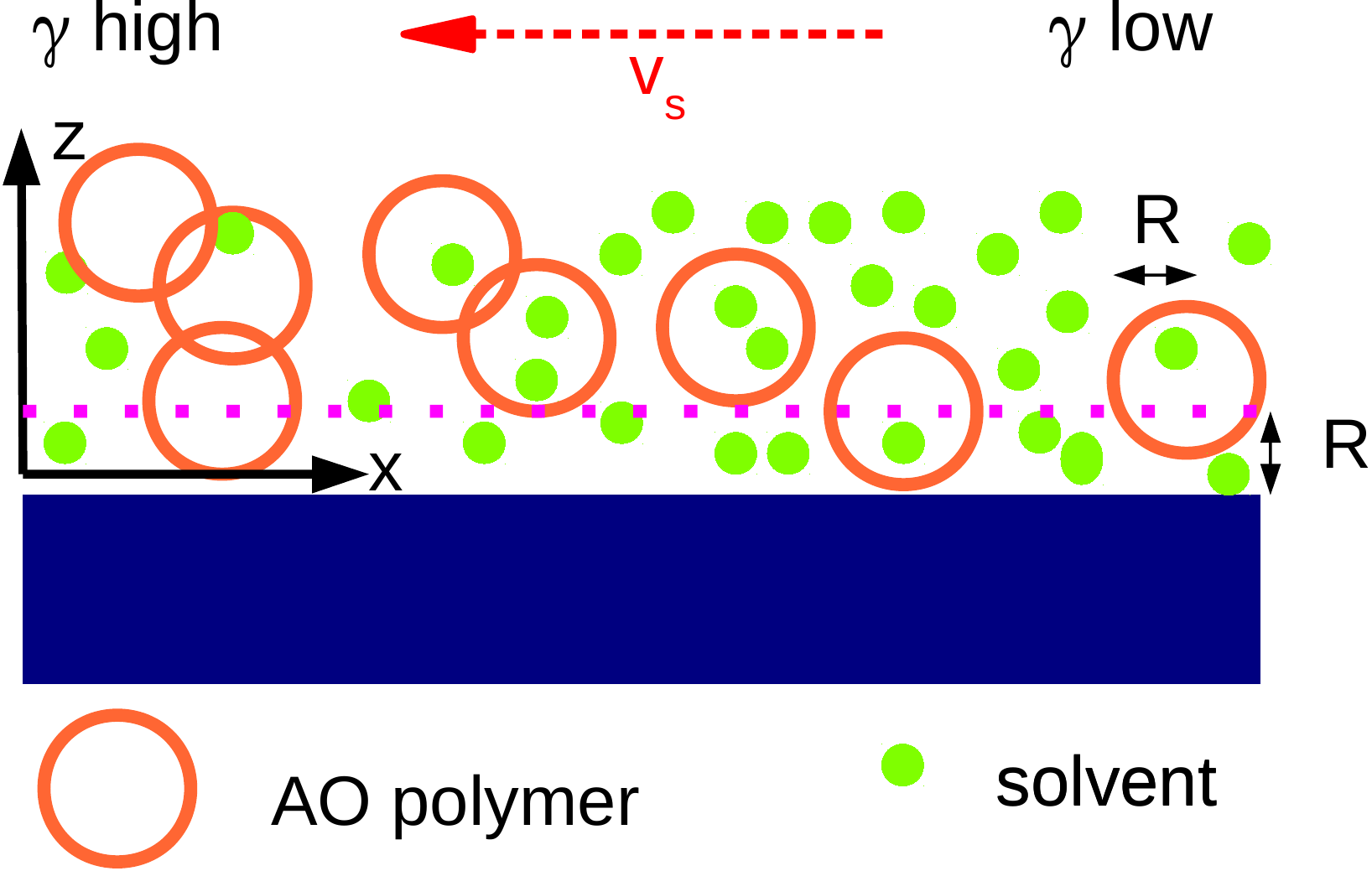}
  \caption{A schematic of a flat wall (blue) in contact with AO-model
    polymer (orange) in a solvent of smaller molecules (green).  The wall
    excludes the polymer from layer of width $R$, the top of this
    layer is indicated by a dotted line.  There are gradients of both
    the polymer and solvent concentrations, parallel to the wall along
    the $x$ axis. The flat wall is assumed stationary, and then the
    fluid flows to the left.
    $\gamma$ is the wall/solution
    surface free energy.
    \label{ao_schem}}
\end{figure}

As we are in the $\Rc\gg R$ limit, the curvature of the surface of the
colloid can be neglected, and we can thus start by considering the
relative, or slip, velocity $\vs$ between a hard wall (surface of
large colloid) and the polymer solution.  To calculate this we
consider the geometry shown in \Figref{ao_schem}, where the wall and
gradient are along the $x$ axis.  For diffusiophoresis of a solid
particle, the stresses and hence the velocity gradients are localised
to the interfacial region \cite{And89}. Here this is a region of width
$R$.  The standard theory of diffusiophoresis \cite{ALP82, And89}
yields the following expression for the wall slip velocity
\begin{equation}
  \vs=-\frac{\kT}{\eta}\Bigl[\int_0^\infty\!\! z\,(e^{-\beta\varphi(z)}-1)\,dz\Bigr]
  \,\nabla_x\rho_p\,,\label{eq:do1}
\end{equation}
where $\eta$ is the viscosity (assumed equal to the bulk viscosity),
$\varphi(z)$ is the potential between the solute molecules and the
wall, and $\beta=1/\kT$.  This result applies to an ideal solute whose
concentration gradient in the fluid is $\nabla_x\rhop$, and was first
identified by Derjaguin and co-workers in 1947~\cite{derjaguin47}.

For the AO model, the wall potential is just
\begin{equation}
  \varphi(z)=\left\{\begin{array}{ll}
  \infty & z<R\,,\\[3pt]
  0 & z > R\,.
  \end{array}\right.
  \label{ao_wall}
\end{equation}
With this potential, the integral in square brackets in \Eqref{eq:do1}
is just minus the first moment of the width of the exclusion region,
\ie\ $\frac{1}{2}R^2$. Then the slip velocity
\begin{equation}
  \vs=\frac{R^2\kT}{2\eta}\,\nabla_x\rhop
  \label{vsdp}
\end{equation}
(this basic result can be found in Anderson's review
\cite{And89}).  Note that the slip velocity results in flow away from
low concentrations of the polymer and towards higher
concentrations. This can be understood as a wall-bounded Maragoni-like
flow away from where the surface tension is low, and towards where it
is high \cite{ruckenstein81}.  This motion reduces the total
wall/solution surface free energy, as the region of low AO-model polymer
concentration expands.  We show in Appendix \ref{app:gamma}, that the
AO-model polymer contributes $\rhop \kT R$ to the surface tension, and so
the surface tension is highest where the concentration of AO-model polymer
is highest, driving flow of the solution to these high polymer
concentration regions.

\Eqref{eq:do1} rests on the fact that in a mixture relaxing
by diffusion, the hydrostatic pressure $p$ is uniform, so that a
gradient in the osmotic pressure $\Pi$ is balanced by a countergradient
in the solvent contribution to the pressure.  The hydrostatic pressure
is uniform because it is a `fast' variable which relaxes \via\ solvent
flow, and so gradients in $p$ are quickly eliminated.  (We also assume
other gradients, for example, in the temperature are negligible.)  But
the osmotic pressure (and hence counterbalancing solvent contribution
to the pressure) only relax \via\ diffusive motion of the polymer, which
is much slower.  This consideration will be crucial in considering
implicit \versus\ explicit solvent models. Details of the gradients in
the interfacial region are in Appendix \ref{app:slip}.

\subsection{Diffusiophoretic drift of large colloidal particle}
For a particle of size $\Rc\gg R$, curvature of the interface is
negligible. Then, the diffusiophoretic drift velocity of such a particle in
a stationary fluid is $U=-\vs$, where $\vs$ is the above slip velocity
at particle's surface. This apparently trivial result hides a great
deal of subtlety in terms of the underlying low Reynolds number flow
problem \cite{levich_book, ALP82, And89}.  Thus one has
\begin{equation}
  U=\Gamma \,\nabla\ln\phip\,,\quad
  \Gamma= -\frac{\phip\kT}{2\eta R}\sim-\phip\Dp\,.
  \label{dp_vel}
\end{equation}
where we have introduced a diffusiophoretic drift coefficient
$\Gamma$.  We also define $\phip\equiv\rhop R^3$ as the dimensionless
polymer concentration (\ie\ packing fraction).  The final
order-of-magnitude scaling estimate rests on the Stokes-Einstein
expression for the diffusion coefficient of the polymer,
$\Dp\sim\kT/(\eta R)$.

The drift is \emph{away} from the high polymer concentration regions,
\ie\ it tends to cause the large colloid to segregate from the smaller
polymer. This can be understood as being driven by the reduction in
surface free energy when the colloid moves to regions of lower polymer
concentration where the surface tension between the colloid particle
and the polymer solution is lower.

Colloidal particles can also move under the action of gravity (\ie\
in sedimentation). But it is important to note that diffusiophoretic
motion and sedimentation are fundamentally different \cite{And86,
  Bra11}. In sedimentation there is an external force (gravity) acting
on the particle and this causes the falling particle to set up a
long-ranged $1/r$ flow field in the surrounding liquid. By contrast in
diffusiophoresis, the stresses and velocity gradients are largely
localised to the interfacial region (here of thickness $R$), and the
flow outside the immediate interface is much weaker and decays as
$1/r^3$ \cite{And86,And89} .

\subsection{Typical diffusiophoretic drift velocities}
We now estimate how fast are typical diffusiophoretic drift
velocities.  For water at room temperature,
$\kT\approx4\times10^{-21}\,\mathrm{J}$, and $\eta\approx
10^{-3}\,\mathrm{Pa\,s}$. We consider a polymer of radius $R\approx
10\,\mathrm{nm}$, with a maximum concentration $\phip\approx 1$. Then
the diffusiophoretic drift coefficient $\Gamma\approx
-200\,\umsqpersec$.  Thus if the length scale of the gradient
$\lambda_G=100\,\um$, then the drift velocity $U$ is of order
$1\,\um$ per second, and independent of the size of the large
particle \footnote{ This is for a solid particle, for a liquid droplet
  with similar viscosities for the liquids inside and out, the
  diffusiophoretic drift velocity is given by a similar expression to
  \Eqref{vsdp} but with the interfacial width $R$ replaced by the
  particle radius \cite{And89}.  As the particle radius is much
  larger, the diffusiophoretic drift velocity (\ie\ Marangoni effect)
  will also be much larger for a liquid droplet.  In particular we
  predict a non-adsorbing liquid droplet of a low viscosity oil in
  water will move rapidly down a polymer gradient.}.

As noted in recent experimental work \cite{ACY+08, KCR+15, SUS+16},
diffusiophoresis can be considerably more effective than diffusion at
transporting micron-sized colloid particles over large distances.
Rates of transport by directed motion (diffusiophoresis, flow, \etc)
can be compared to rates by diffusion \via\ \Peclet\ numbers.  Here
the \Peclet\ number comparing diffusion to diffusiophoretic motion
for the colloidal particle is
\begin{equation}
\PeCDD=\frac{U\lambda_G}{\Dc}\sim \frac{\phip \Dp}{\Dc}
\end{equation}
where $\Dc$ is the diffusion coefficient of the colloid.  We used the
scaling result for $\Gamma$ in \Eqref{dp_vel} to obtain the final
expression, invoking $U=\Gamma \,\nabla\ln\phip$ and
$\nabla\ln\phip\sim1/\lambda_G$.  Because of this the \Peclet\
number is independent of the length scale of the gradient in the polymer
concentration. For large colloidal particles $\Dc\ll \Dp$ and if
$\phip$ is not too small, diffusiophoretic motion will always be
faster over all length scales (larger than $\Rc$) on which there is a
gradient.

\section{Earlier work on drying films, where an
  implicit solvent was used}\label{sec:ear}
In earlier work that one of us (RPS) was involved in, an osmotic
imbalance mechanism was proposed to explain the motion of large
colloidal particles in a concentration gradient of smaller colloidal
particles \cite{fortini16, fortini17}. This model was used as a possible explanation
for the stratification seen in the drying film
experiments \cite{fortini16, martin16, makepeace17}.  Zhou
\etal\ \cite{zhou17} proposed
what is in effect a simple
dynamic density functional theory (DDFT),
while Howard \etal\ \cite{HNP17a,HNP17b}
proposed a more advanced DDFT;
see Appendix \ref{app:ddft} for a discussion
of DDFT.
In all these models, there is no explicit
solvent. The Langevin dynamics
simulations \cite{fortini16,fortini17,HNP17a,HNP17b}
also use an implicit solvent. There
the
solvent is replaced by friction against a stationary background, plus a
corresponding noise term.
The models \cite{fortini16,fortini17,zhou17,HNP17a,HNP17b} all use the Stokes expression
($6\pi R_c\eta v$,
at velocity $v$)
for the drag in a stationary fluid.
In the experiments there is, of course, a
solvent (water).

The osmotic imbalance mechanism \cite{fortini16, fortini17} argues
that the gradient in the concentration of the smaller colloidal
species gives rise to an imbalance in the osmotic pressure across the
diameter of the large particles, leading to a drift velocity of the
larger colloid.  (The same argument would apply for a colloid particle
in a polymer solution.) The work of Zhou \etal\ \cite{zhou17} and of
Howard \etal\ \cite{HNP17a,HNP17b} is similar in the sense that they too have
models for stratification, in which there is an implicit
solvent. Earlier work \cite{routh13review} on colloids in evaporating
films has also used an implicit solvent.

In \Refcite{fortini16} the size of the effect was estimated as
follows.  The osmotic pressure difference across the particle diameter
is of the order $\Rc\, \nabla(\rhop\kT)$.  This gives rise to a force
\begin{equation}
F\sim -\Rc^3 \, \nabla(\rhop\kT)
\label{osmp}
\end{equation}
since the area over which the osmotic pressure difference acts is of
order $\Rc^2$.  We see that this is essentially equal to the volume of
the large particle multiplied by the osmotic pressure gradient, as in
the generalised Archimedes principle which applies in sedimentation
equilibrium \cite{PBS+12}. \Refcite{fortini16} then argued that the
force $F$ leads to a drift velocity $U$ in accordance with the Stokes
mobility, $U=F/(6\pi\eta \Rc)$.  Using \Eqref{osmp} this predicts a
velocity
\begin{equation}
U \sim -\Bigl(\frac{\Rc}{R}\Bigr)^2\times
  \frac{\phip\kT}{\eta R}\times\nabla\ln\rhop
\qquad\text{[incorrect]}
\label{eq:bad}
\end{equation}
where we emphasise that we now regard this result as containing an
incorrect scaling with particle size.  Compared to the correct
diffusiophoretic drift velocity in \Eqref{dp_vel}, the result above is
a factor $(\Rc/R)^2$ larger, and would overwhelm the former for
$\Rc/R\gg 1$. 

However, both this equation and the Langevin dynamics simulations
neglect the solvent dynamics.  We now discuss why this is incorrect,
and in particular why \emph{solvent backflow} critically modifies the
above result.  This observation is not new, and indeed J\"{u}licher
and Prost make the same point in a different context
\cite{julicher09}. The complete story can be found in a
\tourdeforce\ analysis by Brady \cite{Bra11}.  Our exposition takes a
different, more informal approach, but we believe the essential point
is the same.

The key point is that the force acting on the particle
should additionally include a contribution from the solvent
pressure gradient, as well as the osmotic pressure gradient from the
solute.  The solvent pressure gradient arises from solvent backflow,
which must always be present in a counter-diffusing
solute-plus-solvent mixture (\ie\ relaxing by collective or mutual
diffusion).  Then, it is relatively simple to demonstrate that the
solvent force apparently perfectly cancels the osmotic imbalance
force.

We first recall the fundamental definition of the osmotic pressure
$\Pi$, as the difference between the actual (hydrostatic) pressure $p$
in the system of interest, and the pressure $p_s(\mu_s)$ in a system
comprising pure solvent at the same chemical potential
\cite{thermo_book, atkins_book}, \viz\
\begin{equation}
  p = \Pi + p_s(\mu_s)\,.\label{eq:psum}
\end{equation}
When there is a gradient in the concentration of a colloidal or
polymer species there will be a gradient in the osmotic pressure
$\Pi$. However, in the absence of an external body force (like
gravity, in a sedimentation equilibrium), the hydrostatic pressure
rapidly relaxes by bulk flow to become uniform ($\nabla p=0$), on a
time scale which is much faster than colloidal particles or polymers
can diffuse to eliminate a gradient in the osmotic pressure.  If $p$
is uniform, we have that the osmotic pressure and the solvent pressure
have equal and opposite gradients ($\nabla\Pi=-\nabla p_s$).  Applying
the above osmotic imbalance argument we conclude that the force
arising from the solvent pressure gradient cancels that arising from
the osmotic pressure gradient, or in other words the net force
$\Rc^3\,\nabla p=0$ (since $\nabla p=0$).

At this point we have apparently argued ourselves into a corner: if
there is no net force, there can be no drift, and no
stratification mechanism.  However such a conclusion is incorrect, as
the theory of diffusiophoresis shows.  Whilst there is indeed \emph{no
net force}, there is a \emph{force dipole} at the wall, arising
from the solute structuring by the wall potential.  In the present
case (AO model) this can be turned into a quantitative argument which
recovers the classical diffusiophoretic drift velocity, highlighting
that it is in fact the gradient in the solvent pressure, in the
exclusion layer adjacent to the particle, that causes the drift in the
AO model case.  This is explained in Appendix \ref{app:slip}.

To complete the story we return briefly to sedimentation equilibrium.
In this case, it is the solvent pressure $p_s(\mu_s)$ that must be
constant, not the hydrostatic pressure $p$.  The former is constant
because in equilibrium the solvent chemical potential $\mu_s$ is
everywhere the same; the latter (hydrostatic pressure) is not constant
because there is a body force acting on the fluid as a whole, due to
gravity.  Invoking \Eqref{eq:psum} once more, we see that $\nabla
p=\nabla \Pi$ in this case, and thus the integrated effect of the
hydrostatic pressure gradient can be accounted to an osmotic pressure
gradient, which leads to a force on the particle.  Balancing this
force against gravity recovers the generalised Archimedes principle
\cite{PBS+12}.  Thus in sedimentation equilibrium it is possible to
ignore the solvent degrees of freedom.

In implicit solvent models, such as the Langevin dynamics simulations
and the DDFT models
used to describe film drying \cite{HNP17a, HNP17b,zhou17}, solvent backflow
is ignored. 
Zhou {\it et al.}'s \cite{zhou17}
model is effectively a simple DDFT.
The result is of course that the effect of $\nabla p_s$
is thrown out, leaving the osmotic pressure imbalance force
$\Rc^3\,\nabla\Pi$.  Put another way, if we simplify a model by using an
implicit solvent that is defined to be uniform (and indeed static),
then unbalanced gradients in the osmotic pressure appear as though
they arise from an external force such as gravity.  As we have just
argued, this mistake leads to a grossly overestimated drift velocity,
with the `wrong' size scaling, as in \Eqref{eq:bad}. Therefore,
implicit-solvent models should be used with caution when applied to
dynamical situations, including interpreting experimental data on
drying films.

Although the above story is complete, there is an alternative
viewpoint that is worth touching upon.  As Brady has emphasised
\cite{Bra11}, another way to think about the problem is to retain an
explicit representation of the solute particles (polymers, in the AO
model), and incorporate the correct physics by properly accounting for
hydrodynamic interactions.  From this viewpoint, there is indeed an
imbalanced osmotic pressure gradient force on a colloid particle, and
this indeed leads to a `primary' drift velocity in accord with the
Stokes mobility.  However, this primary effect is
almost completely cancelled by the long-range hydrodynamic
interactions, which account for the solvent backflow.

Returning momentarily to the DDFT modeling, we show in Appendix
\ref{app:ddft} that this formalism gives rise to chemical potential
gradients which can essentially be ascribed to unbalanced osmotic
pressure gradients.  Since the underlying density functional theories
are known to be very accurate \cite{SLB+00, Rot10}, it would be nice
to find a way to incorporate this into the modeling.  The `missing
link' appears to be in the specification of the Onsager coefficients
which relate the chemical potential gradients to the fluxes (see Appendix
\ref{app:ddft}).  In the recent work \cite{zhou17, HNP17a, HNP17b} the
Onsager coefficients were chosen to correspond to independent Stokes
mobilities (\ie\ friction against a stationary background), neglecting
hydrodynamic interactions as an obvious simplification.  With
hindsight, it is clear that much more careful attention needs to be
paid to this aspect.  We leave this for future work.

\section{Application of Asakura-Oosawa model to film drying}\label{sec:dry}
We now turn to the application of the theory in Section \ref{sec:dp} to
the problem of stratification of colloid particles in a drying polymer
solution, first considering the sub-problem of the distribution of the
polymer.

\subsection{Distribution of Asakura-Oosawa model polymer in a drying film}
The presence or absence of gradients in the concentration of the
polymer, is determined by a competition between the evaporation
speed $\vev$, and diffusion of the polymer.  This is quantified by
the \Peclet\ number \cite{trueman12, routh13review, fortini16,
  fortini17} for the AO-model polymer molecules in the drying film,
$\Pefilm$, which compares the time scale for diffusion to that for
drying of the film.  This is defined by
\begin{equation}
\Pefilm=\frac{\vev H}{\Dp}
\end{equation}
where $H$ is the film thickness.  If $\Pefilm<1$ drying is slower
than diffusion. Diffusion smooths out concentration gradients, and so
here there are only weak gradients.  But if $\Pefilm\agt O(1)$,
diffusion cannot keep up with accumulation below the descending
interface. Hence gradients will form and diffusiophoresis will occur.

The analysis is facilitated by a happy coincidence: since the polymer
molecules in the AO model are by definition ideal, we can use the
exact solution for a diffusing ideal gas in front of a moving
impermeable wall to represent the effect of the drying solvent front.
This leads us to consider a diffusing ideal gas between two
impenetrable walls. A diffusing ideal gas satisfies
\begin{equation}
\frac{\partial \phip}{\partial t}=\Dp\frac{\partial^2 \phip}{\partial z^2}
\label{aodiff_film}
\end{equation}
with the $z$ axis normal to the walls.  The boundary conditions are
the two walls, plus the initial condition that at time $t=0$, we have
a uniform distribution of the polymer, at an initial packing fraction
$\phipinit$.

The bottom wall is fixed at $z=0$. This models the substrate the
film is on.  The boundary condition here is just zero flux.  The top
wall is the solvent/air interface and is descending at speed $\vev$.
It therefore has a position
\begin{equation}
  \zint(t)=H-\vev t=(1-\tstar)H
\end{equation}
at time $t$, where we have defined the reduced time
\begin{equation}
  \tstar=\frac{\vev t}{H}\quad ({}\le 1)\,.
\end{equation}
Here the boundary condition is
\begin{equation}
\Dp\left(\frac{1}{\phip}\frac{\partial \phip}{\partial z}\right)_{\zint,t}
=-\vev
\label{bczint}
\end{equation}
which is just the criterion for a diffusing ideal gas that the particles
actually at the descending interface must be descending at the same speed as
the interface.

In general we must solve for $\phi(z, t)$ numerically. However in the
limit we are mostly interested in, where $\Pefilm\gg 1$, there is
range of times $\tstar$ where a gradient is established but where the
gradient is far from the bottom surface, \ie\ where the accumulation
zone beneath the descending interface has not yet reached the
bottom. Then the profile can be obtained by taking the $H\to\infty$
limit and using the exact solution in that limit of Fedorchenko and
Chernov \cite{fedorchenko03}, see Appendix \ref{app:dry}.

\begin{figure}
  \includegraphics[width=7cm]{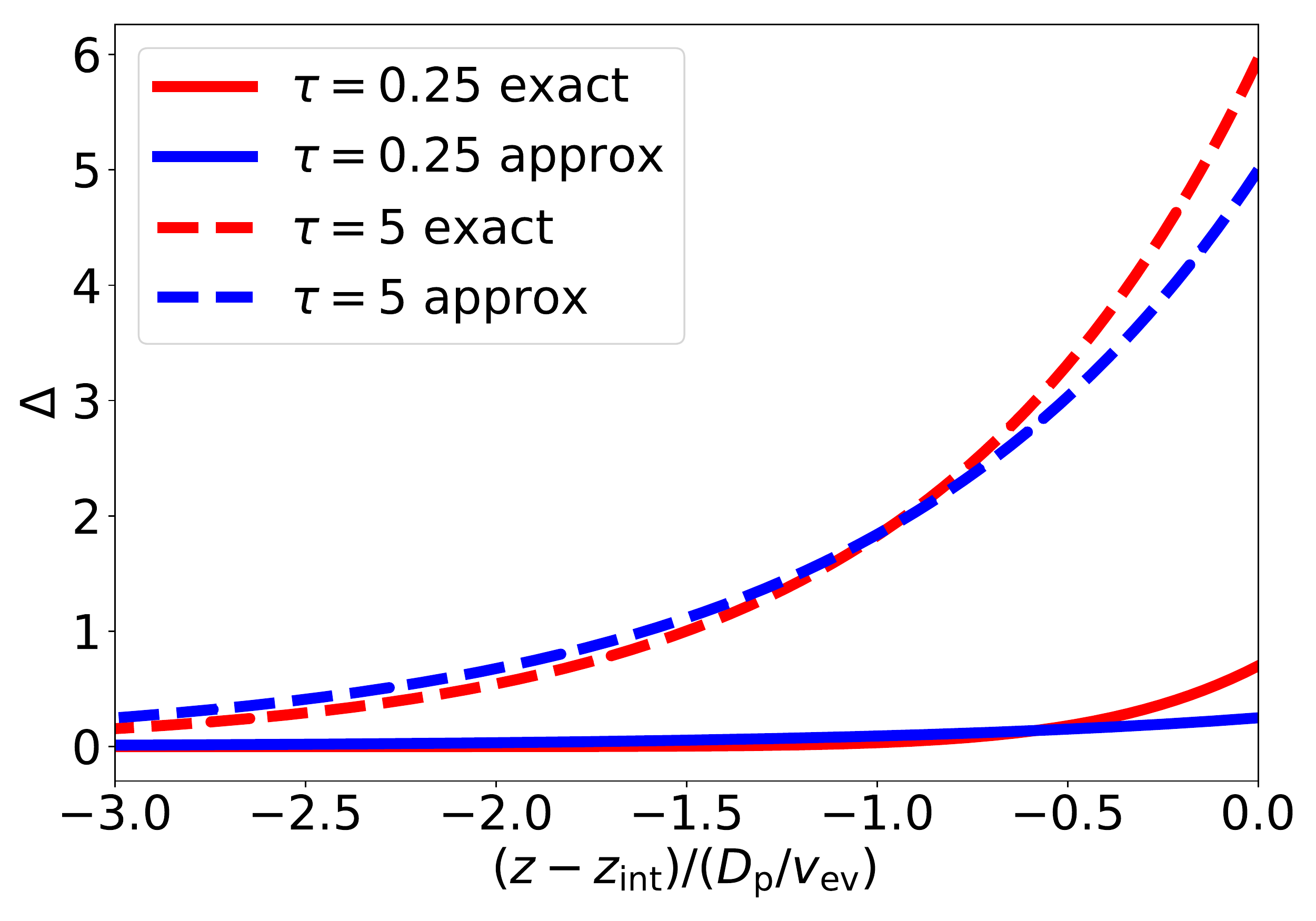}
  \caption{Plots of excess AO-model polymer $\Delta(z,\tau)$ accumulating in
    front of a wall moving at speed $\vev$ (see Appendix
    \ref{app:slip}).  The red curves are the exact solution of
    Fedorchenko and Chernov \cite{fedorchenko03}, while the blue
    curves are the approximate solution in \Eqref{phip_approx}. The
    solid curves are at the early time $\tau=t\vev^2/\Dp=0.25$ when
    the gradient is being established, while the dashed curves are at
    the later time $\tau=5$, when the exponential $z$ dependence is
    well established. At still later times, the height of the
    exponential continues to increase linearly, while the width
    remains constant.\label{fedor}}
\end{figure}

The full solution of Fedorchenko and 
Chernov \cite{fedorchenko03, poon13} is a little complicated,
but after short time $\tstar=1/\Pefilm$, an accummulation
zone is established
of constant width $D/\vev$, and linearly increasing height.
In that regime ($\tstar\Pefilm \gg 1$), the solution simplifies to
\begin{equation} 
\phip(z, t)\approx\phipinit \left(1+
\Pefilm \tstar\exp\left[-\frac{|z-\zint|}{\Dp/\vev}\right]\right)\,.
\label{phip_approx}
\end{equation}
We have plotted profiles in \Figref{fedor}.

The gradient in $\phip$ in this regime ($\tstar\Pefilm \gg 1$) is
\begin{equation} 
\frac{\partial\phip(z, t)}{\partial z}\approx
\frac{\vev\phipinit\Pefilm \tstar}{\Dp}
\exp\left[-\frac{|z-\zint|}{\Dp/\vev}\right]\,.
\label{film_grad}
\end{equation}
Both $\phip$ and its gradient are a maximum at the interface.

\subsection{Diffusiophoresis in a drying film}
We now want to calculate the effect of diffusiophoresis on colloidal
particles in a drying film. To do this we simply combine the results
for diffusiophoresis in AO-model polymer gradient, \Eqref{dp_vel},
with \Eqref{film_grad} for the gradient in a drying film.
This gives the diffusiophoretic drift velocity
\begin{equation}
U(z, t)=-\vev\Pefilm\phipinit \tstar
\exp\left[-\frac{|z-\zint|}{\Dp/\vev}\right]\,.
\label{film_dpvel}
\end{equation}
The dynamics of the colloidal particles during drying of the film are
determined by the competition between two velocities: $U$ and
$\vev$. To begin with we keep things as simple as possible and neglect
diffusion (the $\Dc\to 0$ limit).  In this limit a particle below the
interface moves down with speed $U(z, t)$, while actually at the
interface the descending interface velocity $-\vev$ forms an upper
bound to the particle velocity. In other words, at the interface
particles must descend at least as fast as the interface, but can go
faster. Thus in a drying film the velocity of a colloidal particle is
\begin{equation}
\vc(z, t)\sim\left\{\begin{array}{ll}
\min\left(-\vev , -\vev\Pefilm\phipinit \tstar\right)\,, & (z = \zint)\\[3pt]
U(z, t)\,. & (z< \zint) 
\end{array}\right.
\label{v_phor_film}
\end{equation}
If we look at the behaviour actually at the interface, we have
\begin{equation}
\vc(\zint,t)\sim\left\{\begin{array}{ll}
-\vev\,, & (\Pefilm\phipinit\tstar\le 1) \\[3pt]
-\vev\Pefilm\phipinit \tstar\,. & (\Pefilm\phipinit\tstar> 1)
\end{array}\right.
\end{equation}
There are thus two regimes, the first is for $|U|<\vev$, where the
particles are swept up by the descending interface, forming
a layer there.  The second is for $|U|>\vev$, where
diffusiophoretic motion is faster than the speed of descent of the
interface. Then the colloidal particles outrun the descending
interface, and no particles accumulate at the interface. In this
latter case the colloidal particles are depleted from a layer of
thickness of order $\Dp/\vev$ below the interface, this is the
thickness of the accumulated layer of polymer.

\begin{figure}
  \includegraphics[width=8cm]{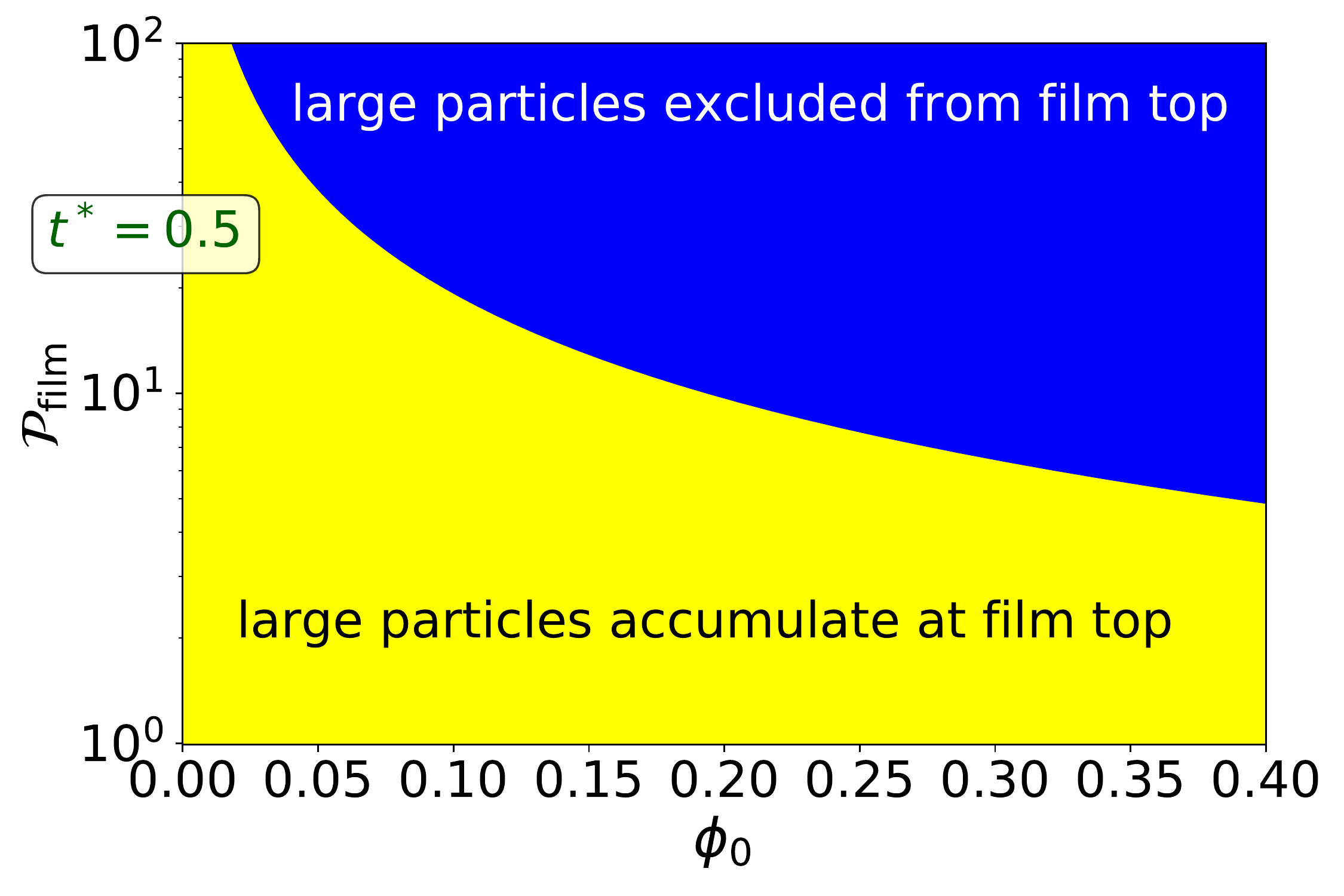}
  \caption{Plot of the $(\Pefilm, \phipinit)$-plane, 
    separated into the region where the large particles accumulate at
    the descending interface at the top of the drying film (yellow),
    and the region where the large particles are excluded from the
    region at the top of drying film just below the interface (blue).
    Large values of $\Pefilm$ and $\phipinit$ give rise to strong
    diffusiophoresis and drive particles away from the top of the
    film.\label{phi_peclet}}
\end{figure}

\subsection{When $\Pefilm$ and $\phipinit$ are large,
stratification occurs with depletion of colloidal particles
from a top layer}
As the reduced time $\tstar\le 1$, then unless the product
$\Pefilm\phipinit>1$ depletion of the particles from the top surface
will not occur during drying.  In \Figref{phi_peclet} we show the
$(\Pefilm, \phipinit)$-plane, and have shaded in blue the region
where depletion has started by the time $\tstar$ reaches $1/2$. This
figure follows a similar plot made by Zhou \etal\ \cite{zhou17} for
their model.  The curve separating the regions of accumulation and
depletion is
\begin{equation}
  \Pefilm = 1/\left(\phipinit \tstar\right)\quad
  \Bigl({}=2/\phipinit \>\>\text{for}\>\> \tstar=1/2\Bigr)\,.
\label{strat}
\end{equation}
Note that there is always accumulation of the colloidal
particles at early times as then
there are no gradients. Gradients take time to build up. But at
high enough polymer concentrations and \Peclet\ numbers,
this turns to depletion of the colloidal particles
from the top layer, later on during drying.

\begin{figure}
  \includegraphics[width=8cm]{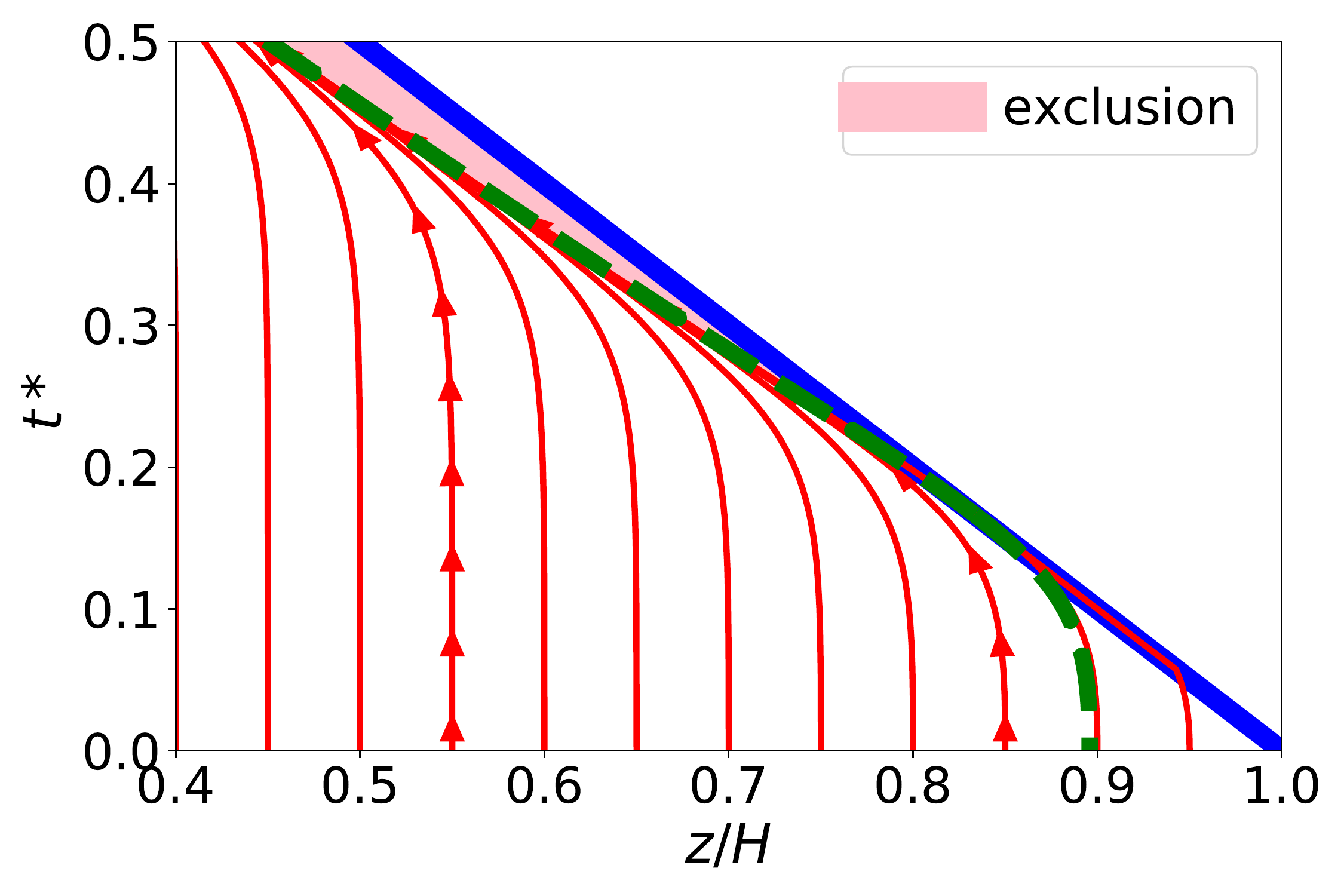}
  \caption{Trajectories $z(\tstar)$ of a large particle as a function
    of reduced time $\tstar$ (red curves).  Two of the trajectories
    have arrows to indicate the direction of the movement.  The
    position of the top interface, $\zint$, is shown in blue.  The
    green dashed curve is the separatrix that divides trajectories
    that touch the top interface from those that do not.  Colloidal
    particles are excluded from the triangular region between the green dashed
    curve and the interface. This exclusion zone is
    shaded in pink.  Calculations are for $\Pefilm=20$ and
    $\phipinit=0.3$.\label{charac}}
\end{figure}

In the absence of diffusion, the velocity $\vc(z, t)$ in
\Eqref{v_phor_film} completely describes the time evolution of the
positions of the particles during drying.  The height of a particle at
time $t$ is simply obtained by integrating ${\rm d}z(t)/{\rm
  d}t=v_{\rm C}$. Using this we further develop our understanding of
the behaviour of the large particles in a drying film by considering
the fate of a set of particles that at $t=0$, are uniformly spaced across the
height of the film. The results are shown as the red curves in
\Figref{charac}, where each red curve followed upwards traces out
the trajectory of a particle. The blue line is the position of the top
interface.  Trajectories that start near the bottom of the film
(\ie\ left hand side of plot) are relatively unaffected and remain
close to vertical, those in the middle of the film are pushed to the
left away from the interface, while those near the top hit the
descending interface for a time, and then later are pushed away from
the top (assuming they can escape the interface).

Note that particles that start near the top of the initial wet film
are caught by the descending interface early on in drying. But then
later as the polymer gradient increases, these particles
are pushed away from the descending interface.  For the parameters
used in \Figref{charac}, $U$ at the surface exceeds $\vev$ at
$\tstar\approx0.17$ (being the point where the green dashed separatrix
touches the interface). At later times there is an expanding zone
beneath the descending interface where $|U|>\vev$ and so the colloidal
particles are pushed down to further away from the interface.  This
exclusion region is shown in pink in \Figref{charac} and there are no
colloidal particles in it, but the concentration of the AO-model polymer is
highest there. In this sense the AO-model polymer and the particles have
\emph{kinetically segregated} in the drying film, with the smaller AO-model
polymer on top.  This is as found in earlier work on mixtures of small
and large colloidal particles \cite{fortini16, fortini17, martin16,
  HNP17a, zhou17, makepeace17}.

Unlike in earlier modelling work we have taken solvent flow into
account. However, our model polymer is simple, and we have not
considered colloid-colloid interactions, so our predictions may not apply
for high colloid packing fractions.  The width of the layer without
colloidal particles is set by the width of the accumulation zone of
the AO-model polymer, which is $\Dp/\vev$. For polymer of radius
$R=10\,\mathrm{nm}$ in water, $\Dp\approx 20\,\umsqpersec$ and so at an
evaporation rate of $\vev=0.1\,\umpersec$ \cite{utgenannt16,
  Ekanayake:2009bs}, this gives a top layer $200\,\um$ thick. This
evaporation rate is typical for room temperature evaporation of water
in still air, heating \cite{utgenannt16} can increase this by a factor
of 10, which decreases the top layer thickness to $20\,\um$.

It is worth noting that as typical film thicknesses are hundreds of
microns \cite{Keddie:2010ta, Ekanayake:2009bs}, and evaporation rates
range from 0.1--$1\,\umpersec$, that for polymers (or particles)
$10\,\mathrm{nm}$ across, film \Peclet\ numbers will be in the range
1--100. Film \Peclet\ numbers for larger colloidal particles will then
be in the range 10--$10^4$. However, for small molecules (\eg\ ions,
co-solvents) we expect film \Peclet\ numbers $<O(1)$ under most
circumstances. Thus diffusiophoresis due to gradients in small
molecules will typically be small, as there will be no gradients to
drive them.  Our \Eqref{strat} encompasses this, as it predicts
stratification occurs only for large \Peclet\ numbers, which
corresponds to large $R$ because $\Pefilm\sim 1/D\sim R$ (with a
Stokes-Einstein diffusion constant).

We note that significant concentration gradients of small
molecules or ions can be produced by surface evaporation in saturated
porous media \cite{veran-tissoires12}.  This suggests interesting
diffusiophoretic phenomenon may be observable in these situations, and
indeed the partner phenomenon of diffusio-osmosis may be relevant
(\ie\ pore-scale flow produced by a wall slip velocity, as in
\Eqref{eq:do1}).

We have not considered diffusion of the colloidal particles, but we
expect this to only perturb our results. To observe diffusiophoretic
effects we require for the polymer that $\Pefilm>1$.  As the
colloidal particle is larger than the polymer, the relevant
\Peclet\ number for the colloid particle is $\Pefilm'=\vev
H/\Dc\gg 1$.  Therefore, during drying, the
colloidal particle will only be able to diffuse short distances, the ratio of the
distance diffused to $H$ is
$({\tstar/\Pefilm'})^{1/2}\ll 1$.

\section{Conclusion}\label{sec:conc}
Drying of a liquid film is an inherently out-of-equilibrium
process. It will always lead to gradients, and these gradients can
drive diffusiophoresis. To explore this we studied a very simple model
system: diffusiophoretic drift of large colloidal particles in
response to a gradient in polymer molecules. We used the
Asakura-Oosawa polymer model.  Due to its simplicity, we obtained a very simple
analytic expression for the diffusiophoretic drift velocity, our
\Eqref{dp_vel}. Also due to its simplicity, the physical mechanism
is clear: the polymer increases the surface tension of the interface
between the colloid and the polymer solution.  This drives motion of
the particle towards lower polymer concentrations, as in a
wall-bounded Marangoni effect.  The velocity gradients associated with
this motion are localised to the interface.
This interface extends out from the particle surface to a distance of
order the polymer size. The resulting
diffusiophoretic drift is opposed by viscous dissipation
largely localised to this interfacial region.

When there is diffusiophoresis in a drying liquid films there are two
competing velocities: the evaporation velocity $\vev$, and the
diffusiophoretic drift velocity $U$.  For large colloidal particles, when at
the top interface $\vev>|U|$ the particles accummulate here, but when
$|U|>\vev$ diffusiophoresis pushes particles away from the top
interface.  This creates a layer beneath the descending interface
where there are no colloidal particles.

Since we expect the essential physics to apply also to mixtures of
large and small colloid particles (see below), this provides an
explanation for the experimental results of Fortini
\etal\ \cite{fortini16}, Martin-Fabiani \etal\ \cite{martin16} and
Makepeace \etal\ \cite{makepeace17}.  They studied dried films made
from mixtures of large and small particles, and found under some
conditions that the large particles were excluded from the top of the
dried film, and there were only small particles there.  The films
stratify during drying, with the small particles on top.

Earlier models \cite{fortini16, martin16, zhou17, HNP17a, fortini17}
also predicted this stratification but as discussed in section
\ref{sec:ear} these models neglected the effect of solvent backflow.
This omission does not change the direction of stratification, but it
does significantly overestimate the effects of diffusiophoretic drift.
Neglecting backflow results in a drift velocity that increases as the square of
the radius of the particle. When the solvent backflow is taken into
account, the correct diffusiophoretic drift velocity is independent of
colloid particle size, and much weaker.  Although this is the case, we
have demonstrated in section \ref{sec:dry} that the correct
diffusiophoretic drift velocity still predicts stratification, under
experimentally accessible conditions, and in accord with experimental
observations.

There is growing interest in diffusiophoresis, and not only is the AO
model an especially simple example, polymers have a number of
advantages for engineering controlled diffusiophoretic motion. Their
effect can be tuned by varying their size, and as they are larger than
ions, they diffuse more slowly, making it easier to establish the
concentration gradients needed.

In this study we have chosen the AO model due to its simplicity, but
the experiments on drying films use near-hard-sphere mixtures
\cite{fortini16, martin16, makepeace17}.  The diffusiophoretic drift
coefficient $\Gamma$ scales as $\gamma/\eta$ (where note $\eta$ 
should be the viscosity in
the structured layer adjacent to
the large colloid surface). Hard sphere interactions increase both
$\gamma$ \cite{heni99} and (bulk) $\eta$ \cite{meeker97}, but neither
effect is dramatic at volume fractions up to around 20--30\%. So, at
all but high packing fractions, we expect the AO model to be roughly
quantitatively predictive of the behaviour of colloidal suspensions at
comparable packing fractions.  The approximations involved in this
assumption can be addressed in part by using more accurate models of
hard sphere suspensions, such as are available from modern density
functional theory (DFT).  The resulting dynamic DFT models are
very appealing \cite{HNP17a, HNP17b}, but it is clear, as we have
identified at the end of section \ref{sec:ear}, that work needs to be
done to address the neglect of solvent backflow effects in this
approach.

\begin{acknowledgments}
PBW would like to thank Sangwoo Shin and Howard Stone for numerous illuminating discussions around diffusiophoresis, and RPS would like
to thank Joseph Keddie for many helpful conversations.
\end{acknowledgments}

\appendix

\section{Wall surface tension in the Asakura-Oosawa model}\label{app:gamma}
Here we calculate the surface tension $\gamma$ in the AO model, for a
polymer solution against a non-adsorbing hard wall.  The surface
tension is the excess (over bulk) of the grand potential per unit area
of the surface
\begin{equation}
  \gamma=\gamma_0+{\textstyle\int_0^\infty} (\omega(z)-\omega_b)\,dz
\end{equation}
where $\gamma_0$ is the surface tension of pure solvent, $\omega(z)$
is the grand potential density at a height $z$ above the wall, and
$\omega_b$ is the bulk grand potential density.

As the AO model is ideal, the grand potential density
$\omega[\rho(z)]$ is just the ideal term plus the interaction with the
wall
\begin{equation}
  \omega(z)=\rho\kT[\ln({\rho}/{\rho_b})-1]+\rho\varphi
\end{equation}
where $\rho_b$ is the bulk polymer
density.  The variational principle $\delta\omega/\delta\rho(z)=0$ is
satisfied by the Boltzmann distribution,
$\rho(z)=\rho_b\exp(-\beta\varphi)$. So, we have
\begin{equation}
  \gamma=\gamma_0-\rho_b\kT{\textstyle\int_0^\infty}(e^{-\beta\varphi}-1)\,dz\,.
\end{equation}
In the AO model this is easy to evaluate, and we get
\begin{equation}
  \gamma=\gamma_0+\rho_b\kT R\,.
  \label{ao_gamma}
\end{equation}
Thus non-adsorbing polymers \emph{increase} the surface tension by an
amount equal to the osmotic pressure $\rho_b\kT$, multiplied by the depletion layer
thickness. Note that the surface excess is
\begin{equation}
  {\textstyle\int_0^\infty}(\rho-\rho_b)\,dz =
  \rho_b {\textstyle\int_0^\infty}(e^{-\beta\varphi}-1)\,dz\,
=-\rho_b R\,.
\end{equation}
This is \emph{negative}, thus the increased surface tension can be
viewed as a result of the generic Gibbs adsorption isotherm result,
\ie\ $\partial\gamma/\partial\mu = \rho_bR$ where $\mu=\kT\ln\rho_b$.

Diffusiophoretic drift can now be understood as a consquence of a
wall-bounded surface tension gradient \cite{ruckenstein81}.  In
particular we can write $\vs=(\Lstar/\eta)\,\nabla\gamma$, as though the
surface tension gradient (force per unit area) is localised at a
height $\Lstar$ above the actual surface.  
In this simplified picture, the fluid undergoes
uniform shear in the interfacial region $0<z<\Lstar$.  For the AO
model, $\Lstar=\frac{1}{2}R$.  Compared to the actual flow field
solved next, this picture is certainly oversimplified, but 
nevertheless is useful for gaining \adhoc\ insights.

\section{Flow field in interfacial region}\label{app:slip}
To obtain a better understanding of the molecular origins of
diffusiophoretic drift we consider a polymer solution gradient next to
a non-adsorbing wall, as shown in \Figref{ao_schem}, and solve for the
flow field.  In this geometry the velocity is only along
$x$-direction, and is only a function of $z$, \ie\ we have to compute
$v_x(z)$. The relevant component of the Stokes equation is
\begin{equation}
 \eta \frac{{\rm d}^2v_x}{{\rm d}z^2}=\nabla_x p\,.
  \label{stokes}
\end{equation}
In the bulk fluid ($z>R$), the hydrostatic pressure $p$ is uniform,
but in the interfacial region ($z<R$) the AO-model polymer is excluded
and so in our model the relevant pressure is that of the solvent. Thus
although in the bulk there is no gradient in the hydrostatic pressure,
in the interface there is a gradient in the hydrostatic pressure, and
this gradient is equal to the gradient in the solvent pressure in the
bulk. As we have asserted in section \ref{sec:ear}, the solvent
pressure gradient satisfies $\nabla_x
p_s=-\nabla_x\Pi=-\kT\,\nabla_x\rho_b$ where $\nabla_x\rho_b$ is the
polymer concentration gradient in the bulk.  Se we have
\begin{equation}
  \nabla_x p = \left\{\begin{array}{lc}
  -\kT\,\nabla_x\rho_b\,, & z< R\\[3pt]
  0\,. & z \ge R
  \end{array}\right.
  \label{p0int}
\end{equation}
We insert \Eqref{p0int} in \Eqref{stokes}, and integrate twice
applying the boundary conditions $v_x=0$ at $z=0$ and ${\rm d}v_x/{\rm
  d}z\to0$ as $z\to\infty$ (in practice, this holds at $z=R$ in this
model). This yields the velocity profile
\begin{equation}
  v_x(z)=\left\{\begin{array}{lc}
  \left[z(2R-z)\kT/(2\eta)\right]\,\nabla_x\rho_b\,, & z< R\\[3pt]
  \left[R^2\kT/(2\eta)\right]\,\nabla_x\rho_b\,. & z \ge R
  \end{array}\right.
  \label{vx}
\end{equation}
This comprises parabolic (half-Poiseuille) flow in the depletion
layer, continuous with plug flow in the bulk.  The resulting wall slip
velocity $\vs=v_x(\infty)$ (${}=v_x(R)$) is identical to \Eqref{vsdp}.

To anyone familiar with the derivation of the classic result in
\Eqref{eq:do1} this would not be surprising, but it perhaps sheds an
interesting light on the mechanism in the present problem. In words:
the gradient in polymer concentration is necessarily associated with a
counter-gradient in the solvent chemical potential (to maintain
constancy of overall bulk hydrostatic pressure).  This generates a
\emph{real} solvent pressure gradient in the depletion layer adjacent
to the large particle surface.  This pressure gradient drives a thin
film flow, resulting in an effective wall slip velocity on the scale
of the particle.  This leads to diffusiophoretic drift of a suspended
particle.  As noted, this drift is in the
direction of reducing the interfacial free energy of the colloid
particle in the polymer solution.

\section{Fedorchenko and Chernov solution}\label{app:dry}
Fedorchenko and Chernov obtained an analytic solution
\cite{fedorchenko03, poon13} for a diffusing ideal gas in an infinite
system below a descending wall.  We are interested in a thin film not
an infinite ($H\to\infty$ system), however we can apply their result
here, but only for $\Pefilm\gg 1$.

Fedorchenko and Chernov's solution \cite{fedorchenko03, poon13},
applied to our AO-model polymer, can be written in terms of 
\begin{equation} 
\phip(z, t)= 
\phipinit\left[1+\Delta(z, t)\right] 
\label{fc1} 
\end{equation}
for $\phipinit$ the initial uniform concentration of polymer in the
film. Here $\Delta(z, t)$ is given by
\begin{equation}
  \begin{split}
    \Delta(z',\tau)&=
    \left(\frac{\tau}{\pi}\right)^{1/2}\exp\left(-\frac{(z'+\tau)^2}{4\tau}\right)\\[3pt]
    &\quad{}+\frac{1}{2}\left(1-z'+\tau\right)e^{-z'}
    \erfc\left(\frac{z'-\tau}{2\tau^{1/2}}\right)\\[3pt]
    &\qquad{}-
    \frac{1}{2}\erfc\left(\frac{z'+\tau}{2\tau^{1/2}}\right)
  \end{split}
\end{equation}
with $z'=|z-\zint|/(\Dp/\vev)$ and $\tau=t\vev^2/\Dp=\tstar\Pefilm$.
After an initial transient, it simplifies to
\begin{equation} 
  \Delta(z, t)\approx 
  \frac{tv_{ev}^2}{\Dp}\exp\left(-\frac{|z-\zint|}{\Dp/v_{ev}}\right)\,.
  \label{asymp} 
\end{equation}
This holds for $\tau\gg 1$ (\ie\ $\tstar\Pefilm \gg 1$).

Note that this gives an accumulation zone below the interface of width
$\Dp/\vev=H/\Pefilm$. When $\Pefilm\gg 1$, this is much less than the
initial film thickness.  Thus the solution in an infinite system is
very close to that in a thin film, except when $\tstar$ is close to
one, because the accumulation zone ends far above the bottom substrate
at $z=0$.  The time to establish this profile is $\Dp/\vev^2$, or in
reduced units $\tstar=1/\Pefilm$.  Note that at $t=0$ the
concentration is uniform, so there is no gradient.

\section{Dynamic density functional theory}\label{app:ddft}
To see how the problem appears from the point of view of dynamic
density functional theory (DFT), we start with an exact DFT for tracer
(\ie\ dilute) colloids in an ideal polymer solution \cite{Rot10},
\begin{equation}
\begin{split}
\beta f&=
{\textstyle\int}\dvol\,\rho_c(\ln\rho_c-1)
+
{\textstyle\int}\dvol\,\rho_p(\ln\rho_p-1)\\[3pt]
&\qquad{}-{\textstyle\int}\dvol\,\dvol'\,
\rho_c(\rvec)\,\rho_p(\rvec')\,f_{cp}(\rvec-\rvec')\,,
\end{split}\label{eq:f}
\end{equation}
where $\beta=1/\kT$.  The second term in this accounts for the
colloid-polymer interaction, and features the Mayer function
$f_{cp}=e^{-\beta\phi_{cp}}-1$ which we leave general for the time
being. This is essentially also the model proposed by Zhou
\etal\ \cite{zhou17}. From this the colloid chemical potential is
\begin{equation}
\beta\mu_c=\frac{\delta(\beta f)}{\delta\rho_c(\rvec)}=
\ln\rho_c(\rvec)-{\textstyle\int}\dvol'\,\rho_p(\rvec')\,f_{cp}(\rvec-\rvec')\,.
\end{equation}
A similar expression obtains for the polymer chemical
potential. Taking the gradient of $\beta\mu_c$ we find
\begin{equation}
\begin{split}
\beta\,\nabla\mu_c
&={({1}/{\rho_c})}\,\nabla\rho_c-{\textstyle\int}\dvol'\,
\rho_p(\rvec')\,\nabla f_{cp}(\rvec-\rvec')\,,\\[3pt]
&={({1}/{\rho_c})}\,\nabla\rho_c+{\textstyle\int}\dvol'\,
\rho_p(\rvec')\,\nabla'f_{cp}(\rvec-\rvec')\,,\\[3pt]
&={({1}/{\rho_c})}\,\nabla\rho_c-{\textstyle\int}\dvol'\,
\nabla'\rho_p(\rvec')\,f_{cp}(\rvec-\rvec')\,,\\[3pt]
&\approx{({1}/{\rho_c})}\,\nabla\rho_c+\Vcp\,\nabla\rho_p\,,
\end{split}
\end{equation}
where $\Vcp=-{\textstyle\int}\dvol\,f_{cp}(\rvec)$ is the excluded
volume between the colloid and polymer, specialising to the AO model
case for which $f_{cp}(\rvec)=-1$ for $|\rvec|\le R+R_c$ and is zero
otherwise.  In the last step we assume that the polymer
concentration is weakly varying on the scale of the colloid diameter.
In a similar manner we find
$\nabla\mu_p\approx({1}/{\rho_p})\,\nabla\rho_p+\Vcp\,\nabla\rho_c$.

Now consider the matrix of Onsager coefficients which relate chemical
potential gradients to fluxes, $\Jvec_i=L_{ij}\,\nabla\mu_j$.  We
shall suppose that the leading diagonal elements are $L_{cc}=-\rho_c
\Dc/\kT$ and $L_{pp}=-\rho_p\Dp/\kT$, but as an \ansatz\ keep the
leading-order off-diagonal effect $L_{cp}=-\rho_c\rho_p X/\kT$ where
the prefactor $X$ is unknown at this point.  Then the fluxes are given
by
\begin{equation}
\begin{pmatrix}
\Jc\\ 
\Jp
\end{pmatrix}
=-
\begin{pmatrix}
\rho_c\Dc & \rho_c\rho_p X\\
\rho_c\rho_p X & \rho_p\Dp
\end{pmatrix}
\begin{pmatrix}
1/\rho_c & \Vcp\\
\Vcp & 1/\rho_p
\end{pmatrix}
\begin{pmatrix}
\nabla\rho_c\\ 
\nabla\rho_p
\end{pmatrix}
\end{equation}
(cancelling $\beta=1/\kT$ throughout).  
On multiplying through 
\begin{equation}
\begin{pmatrix}
\Jc\\ 
\Jp
\end{pmatrix}
\approx
-\begin{pmatrix}
\Dc & \rho_c(\Vcp\Dc+X)\\
\cdots & \Dp
\end{pmatrix}
\begin{pmatrix}
\nabla\rho_c\\ 
\nabla\rho_p
\end{pmatrix}\,.
\end{equation}
In this we have dropped terms in the diagonal elements which are
$O(\rho_c)$ since we suppose we only have tracer amounts of colloid.
Similarly, the lower left off-diagonal term is irrelevant since it
multiplies $\nabla\rho_c$ and makes a negligible contribution to $\Jp$
under the stated conditions.  Thus we arrive at
$\Jc=-\Dc\,\nabla\rho_c-\rho_c(\Vcp \Dc+X)\,\nabla\rho_p$ and
$\Jp=-\Dp\,\nabla\rho_p$.
The term proportional to $\nabla\rho_p$ in $\Jc$ is also proportional
to $\rho_c$.  We therefore identify it as a drift term, corresponding
to a colloid drift velocity $\Uvec=-(\Vcp \Dc+X)\nabla\rho_p$.
Note that although $X\ne0$ corresponds to a \emph{second order} term
in the Onsager matrix, it has been promoted to a \emph{first order}
correction in the drift velocity, essentially because $\rho_p$ in the
Onsager coefficient cancels $1/\rho_p$ in the chemical potential
gradient.

If $X=0$ the drift velocity can be written
$\Uvec=-(\Dc/\kT)\,\Vcp\,\nabla(\rho_p\kT)$, with the interpretation
(reading right to left) that the osmotic pressure gradient in the
polymer solution generates a buoyancy force as in the generalised
Archimedes principle, which when multiplied by the Stokes mobility
gives the drift velocity.  This is the approximation made by Zhou
\etal\ \cite{zhou17}, and
the low density limit
of Howard \etal\ \cite{HNP17a, HNP17b}'s model, which
now clearly corresponds to the argument made by Fortini
\etal\ \cite{fortini16}.  But our central claim is that this neglects
solvent backflow, and overestimates the true diffusiophoretic drift
velocity.  Therefore $X=0$ is inadmissible.

To recover what we claim to be the correct diffusiophoretic drift, we
therefore expect that the prefactor $X$ should be negative, and large
enough to cancel the leading size dependence in the bare Stokes
result.  One could therefore envisage rather crudely `patching up' the
DDFT by including an off-diagonal Onsager coefficient consistent with
the above arguments.  We leave this for future investigation, though
we note that the much more sophisticated analysis by Brady
\cite{Bra11} identifies the exact way that hydrodynamic interactions
conspire to cancel (most of) the bare Stokes result.


%

\end{document}